\documentclass[12pt]{article}

\usepackage{amsmath}
\usepackage{amssymb}
\usepackage{jeoslike}

\def\be{\begin{eqnarray}}
\def\ee{\end{eqnarray}}

\def\u{{\bf u}}

\begin{document}

 \mytitle{Electric and magnetic dipolar response of Germanium spheres: Interference effects, scattering anisotropy and optical forces}
 \myauthor{R.  G\'omez-Medina$^{1,2,\dag}$, B. Garc\'{\i}a-C\'amara$^{3}$, I. Su\'arez-Lacalle$^{1}$, F. Gonz\'alez$^{3}$, F. Moreno$^{3}$, M. Nieto-Vesperinas$^{2,\S}$, J. J. S\'aenz$^{1,*}$}
\myaddress{
$^{1}$ Departamento de F\'{\i}sica de la Materia Condensada, Universidad Aut\'{o}noma de Madrid, 28049
Madrid, Spain.\\
$^{2}$Instituto de Ciencia de Materiales de Madrid, C.S.I.C., Campus de Cantoblanco, 28049 Madrid, Spain.\\
$^{3}$Departamento de F\'{\i}sica Aplicada, Universidad de Cantabria,  Avda. de los Castros s/n, 39005 Santander, Spain
}

\myemail{$^{\dag}$ r.gomezmedina@uam.es ; $^{\S}$ mnieto@icmm.csic.es ; $^{*}$ juanjo.saenz@uam.es}


We address the scattering cross sections, and their consequences, for submicrometer Germanium spheres. It is shown that there is a wide window in the near infrared where light scattering by these particles is fully described by their induced electric and magnetic dipoles.  In this way, we observe remarkable anisotropic scattering angular distributions, as well as zero forward or backward scattered intensities, which until recently was theoretically demonstrated only for hypothetically postulated  magnetodielectric spheres. Also, interesting new effects of the optical forces exerted on these objects are now obtained.

\keywordline (290.5850) Scattering, particles ; (160.1190)
Anisotropic optical materials ; (160.3820)   Magneto-optical materials; (160.4236)  Nanomaterials;
(230.5750) Resonators;

\section{INTRODUCTION}
Electromagnetic scattering from nanometer-scale objects has long
been a topic of large interest and relevance to fields from
astrophysics or meteorology to biophysics, medicine and material
science \cite{1,2,3,4,Purcell,Draine,Halas,Jain,Day}. In the last few years, small particles
with resonant magnetic properties are being explored as constitutive elements of new metamaterials and devices.
Magnetic effects, however, cannot be easily exploited in the visible or infrared regions  due to intrinsic natural limitations of optical materials and  the quest for magnetic plasmons and magnetic resonant structures at optical frequencies \cite{Engheta,10,Moreno2} has then been mainly focused on metallic structures.
The unavoidable problems of losses and saturation effects inherent to these metamaterials in the optical and near infrared regimes have stimulated the study of high-permittivity particles as their constitutive elements
\cite{Kevin_8,Referee5,Moreno1,Referee1,Referee2,Referee3,Wheeler_2009,Kevin_10,Kevin_11,Kevin_12,Kevin}:  For very large permittivities, small spherical particles present well defined sharp resonances \cite{1,2};  either electric or magnetic resonant responses can then be tuned by choosing the appropriate sphere radius.

In the presence of both electric and magnetic properties, the scattering characteristics of a small object present markedly differences with respect to pure electric or magnetic responses. Even in the simplest case of small or of dipolar scatterers, remarkable scattering effects of magnetodielectric particles were theoretically established by Kerker et al. \cite{Kerker} concerning suppression or minimization of either forward or backward scattering. Intriguing applications in scattering cancellation and cloaking \cite{Leonhardt, PendrySS, Aluennano} and  magneto-optical systems \cite{Lakhtakia,Farias,Braulio,Albadalejo_OE,nanomagma} together with the unusual properties of the optical forces on magnetodielectric particles \cite{opex2010,JOSANEW} have renewed interest in the field.

The striking characteristics of the scattering diagram  of small (Rayleigh) magnetodielectric particles
\cite{Kerker,Miros,GCamara} were obtained assuming  arbitrary values of electric permittivity and magnetic permeability.
Nevertheless, no concrete example of such particles that might present those interesting  properties in the visible or infrared regions had been proposed.
Very recently, it has been shown \cite{Andrey,Aitzol} that submicron silicon spheres present dipolar magnetic and electric responses, characterized by their respective first-order Mie coefficient, in the near infrared, in such a way that either of them can be selected by choosing the illumination wavelength. In a later work, it has also been shown
\cite{JOSANEW,PRANEW} that Si spheres constitute such a previously quested real example of dipolar particle with either electric and/or magnetic response, of consequences both for their emitted intensity and behavior under electromagnetic forces.

These properties should not be restricted to Si particles but should also apply to other dielectric materials with  relatively moderate refraction index. In the present work, we  discuss the effects associated to the interference between electric and magnetic dipoles in germanium spheres. The paper is structured as follows. Based on the exact Mie theory, in Sec. 2 we show that both the extinction cross section and the scattering diagrams of submicron Ge spheres in the infrared region can be well described by dipolar electric and magnetic fields, being quadrupolar and higher order contributions negligible in this frequency range.  Specifically, the scattering diagrams calculated at the generalized Kerker's conditions are shown to be equivalent to those previously reported \cite{Kerker,GCamara} for hypothetical ($\epsilon \ne 1$, $\mu \ne 1$) magnetodielectric particles.
Following previous work regarding the peculiar properties of optical forces at the Kerker's conditions
\cite{JOSANEW}, in Sec. 3
we analyze the consequences of the strong scattering anisotropy on the radiation pressure on Ge particles.

\section{EXTINCTION CROSS SECTIONS AND SCATTERING ANISOTROPY OF SUBMICRON GERMANIUM SPHERES}

Consider a non-absorbing dielectric sphere of radius $a$ and dielectric permittivity $\epsilon_p=m_p^2$ in an otherwise uniform medium
with real relative permittivity $\epsilon$ and refractive index $m=\sqrt{\epsilon_h}$. The magnetic permittivity  of the sphere and the surrounding  medium is assumed to be $\mu=1$.
Under plane wave illumination, and assuming linearly polarized light, the incident wave is described by
 \begin{equation}
 E = E_0 \u_x e^{ikz} e^{-i\omega t} \quad, \quad B = B_0 \u_y e^{ikx} e^{-i\omega t}
 \end{equation}
 where  $k = m \omega/c =  m 2\pi/\lambda$, $\lambda$ being the wavelength in vacuum and $B_0 = \mu_0 H_0 =-(m/c) E_0 $ . The
field scattered by the sphere can be decomposed into a multipole series (the so-called Mie's expansion)
characterized by the $\{a_n\}$ electric and $\{ b_n\}$ magnetic Mie coefficients (being $a_1$ and $b_1$ proportional to the electric and magnetic dipoles, $a_2$ and $b_2$ to the quadrupoles and so on). Mie coefficients are related to the scattering phase-shifts  $\alpha_n$ and $\beta_n$ through \cite{1}
\be
a_n &=& \frac{1}{2}\left(1-e^{-2i\xi_n }\right) = i \sin \xi_n e^{-i\xi_n} \label{an}\\
b_n &=& \frac{1}{2}\left(1-e^{-2i\beta_n }\right) = i \sin \beta_n e^{-i\beta_n}. \label{bn}
\ee
The extinction, $Q_{\mbox{ext}}$, and scattering, $Q_S$, cross sections of a dielectric sphere, expressed in the Mie coefficients, read
 \be
 Q_{\mbox{ext}} &=&  \frac{2\pi}{k^2}\sum_{n=1}^\infty \left(2n+1\right) \mbox{Re}\left\{ a_n + b_n \right\} \\
 Q_S &=& \frac{2\pi}{k^2}  \sum_{n=1}^\infty \left(2n+1\right) \left\{|a_n|^2+|b_n|^2 \right\}
 \ee
In absence of absorption, i.e. for real $m, m_p$,  $Q_{\mbox{ext}}=Q_S$.

In the small particle limit ($x \equiv ka \ll 1$) and large particle permittivities ($m_p/m \gg 1$)
the extinction cross section presents characteristic sharp resonance peaks.
At each resonance,  the extinction cross section is of the order of $\lambda^2$ and it is
independent of the particle size or refractive index \cite{1}.
 Interesting applications of  well defined Mie resonances,
associated to low loss and  large dielectric constants, are  accessible for different materials  at  microwave and terahertz frequencies. However,
as $m_p$ decreases there is an increasing overlap between the cross-section peaks and the resonant character disappears.
Since usually non-absorbing materials present low refractive index
 in the infrared (IR) and visible frequency ranges, Mie resonances of small particles in these regimes have not been considered in detail.
 However, a recent the analysis of the cross-section of submicron dielectric particles \cite{Aitzol} show that well defined resonances can be found for materials with  relative refractive index as low as
 $m_p \sim 3$.

\begin{figure}[htbp]
\begin{center}
\includegraphics[width=15cm]{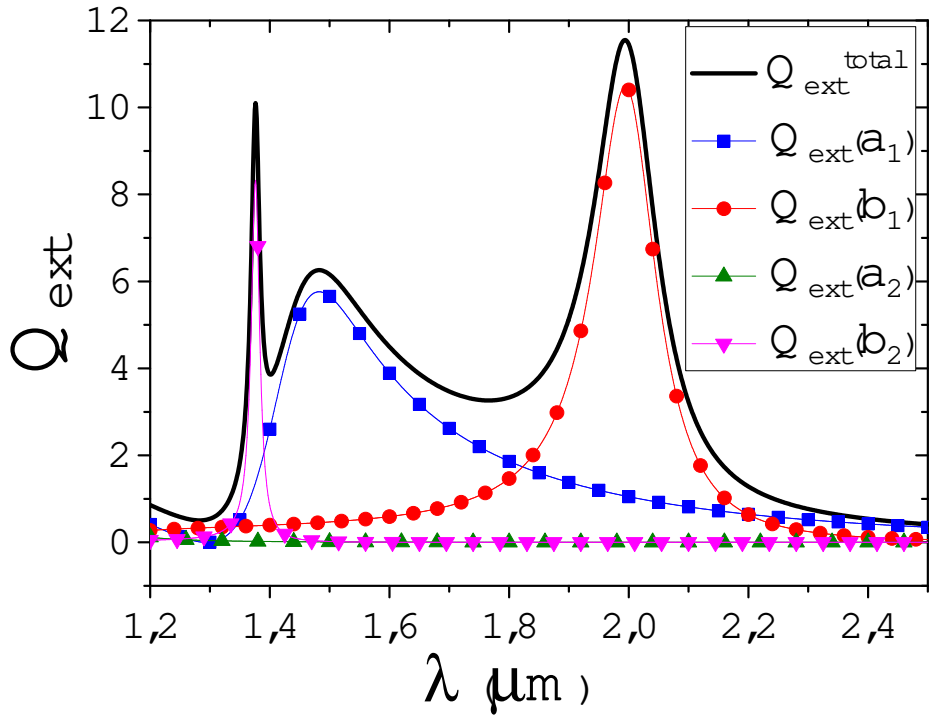}
\caption{Extintion cross-section $Q_{\mbox{ext}}$ versus the wavelength $\lambda$ for a 240nm Ge sphere (the refraction index $m_p=4$ is constant and real in this wavelength range). The contribution of each term in the Mie expansion is also shown. The red line corresponds to the magnetic dipole contribution.}
\label{GeMie}
\end{center}
\end{figure}

 Germanium submicron particles are a good candidate to explore the effects associated to the interference between electric and magnetic responses.  In the micron wavelength regime, within the transparent region of germanium ($\lambda \gtrsim 1.4\mu$m), the refraction index can be well approximated by a real constant $m_p \approx \sqrt{16} = 4$ (see for example \cite{Palik}).  The calculated exact extinction (or scattering) cross section of a 240nm radius Ge sphere in vacuum ($m=1$) is plotted in Fig. \ref{GeMie}.
Although there is an overlap between the first dipolar peaks, the first dipolar magnetic resonance
 (at  $\lambda \approx 2000$nm) and the electric dipolar resonance   (at  $\lambda \approx 1500$nm) are still very well defined. As we will see, this overlap plays a key role in determining the peculiar scattering diagrams of Ge particles.
 Interestingly, for wavelengths larger than $\lambda \approx 1400$nm the cross section is completely determined by the first $b_1$ and $a_1$ coefficients. In other words, in this regime Ge particles can be treated as dipolar particles.

Dipolar particles are usually characterized by their electric and/or magnetic complex polarizabilities, $\alpha_e$ and $\alpha_m$ which may be written in the form \cite{opex2010,Aitzol}
\be
 \alpha_e &=&  3i \epsilon a_1 /(2 k ^3) = \frac{\alpha_e^{(0)}}{ 1- i \frac{2}{3 \epsilon } k ^3
\alpha_e^{(0)}} \\
\alpha_m &=& 3 i b_1 /(2\mu k ^3)=\frac{\alpha_m^{(0)}}{1- i \frac{2}{3} \mu  k ^3
\alpha_m^{(0)}}.
\label{Draine}
\ee
where
 \be
 \alpha_e^{(0)} = - \frac{3 \epsilon }{2 k^3} \tan \xi_1 \quad, \quad \alpha_m^{(0)} = - \frac{3}{2 \mu k^3} \tan \beta_1.
 \ee
 In absence of absorption, $ \alpha_e^{(0)}$ and $ \alpha_m^{(0)}$ are real quantities.

For a pure electric or a pure magnetic dipole, in absence of interferences, the far field radiation pattern is symmetrically distributed between forward and backward scattering. However, when we consider the coherent contribution of both electric and magnetic dipoles, the radiation pattern is mainly distributed in the forward or backward region
according to whether $\Re(\alpha_{e} \alpha_{m}^{*})$ is positive or
negative, respectively \cite{opex2010,JOSANEW}. Interestingly, at the so-called Kerker conditions, $\left| \epsilon^{-1} \alpha_e\right|^2 = \left|\mu \alpha_m\right|^2$, the scattered intensity should be independent of the incident polarization angle:
\begin{eqnarray}
 \frac{dQ_S}{d\Omega} (\theta)=  k ^4
\left| \epsilon^{-1} \alpha_e\right|^2 \left( 1 +  \cos^2 \theta \right) +   2 k^4  \frac{\mu}{\epsilon} \Re(\alpha_e \alpha_m^* ) \cos \theta .
\label{difcross2}
\end{eqnarray}
The interference between electric and magnetic dipoles lead to a number of interesting effects:

{\em i) The intensity in the backscattering direction
can be exactly zero} whenever
\be
\epsilon^{-1} \alpha_e = \mu \alpha_m \quad  ; \quad \frac{dQ_S}{d\Omega}(180^\circ) = 0 \label{firstKK}.
\ee
This anomaly corresponds to the so-called {\em first Kerker condition} \cite{Kerker},
theoretically predicted for magnetodielectric particles having
$\epsilon_p = \mu_p$.

{\em ii)} Although  the intensity can not be exactly zero in the forward direction (causality imposes $\Im\{\alpha_e\}, \Im\{\alpha_m\} > 0$),  {\em  in absence of particle absorption, the forward intensity presents a minimum} at \cite{JOSANEW}
\be
\Re\{\epsilon^{-1} \alpha_e \}= -\Re\{\mu \alpha_m\}  \quad &\mbox{ and }&  \quad  \Im\{\epsilon^{-1} \alpha_e \} = \Im\{\mu \alpha_m\} \nonumber
\\    \frac{dQ_S}{d\Omega}(0^\circ)   &=&
 \frac{16}{9} k^{10} \left|\epsilon^{-1} \alpha_e\right|^4
\label{secondKK}\ee

For lossless magnetodielectric Rayleigh particles, this happens when \cite{Kerker,GCamara} $\epsilon_p = -(\mu_p-4)/(2\mu_p+1)$ and it is known as the {\em second Kerker condition}. Although the original derivation \cite{Kerker} was obtained in the quasi-static approximation ($\Im\{\alpha\} \approx 0$, thus leading  to $dQ(0^\circ)/d\Omega = 0$), the actual intensity for a very small particle goes as $\sim (ka)^{10}$ which, for Rayleigh particles, would be negligible \cite{Aluennano}. However, the  derivation of the unusual scattering conditions (\ref{firstKK})
and (\ref{secondKK}) was obtained \cite{JOSANEW} with the unique assumption that
the radiation fields are well described by dipolar electric and
magnetic fields. This goes well beyond  the Rayleigh limit and
should apply to any dipolar particle. The first line of Eq.
\ref{secondKK} is considered as a {\em generalized Kerker
condition}. Specifically, {\it
the two Kerker conditions also apply to purely dielectric spheres
($\mu_p =1$) providing that their scattering properties may be
described by the two first terms in the Mie expansion}.

\begin{figure}[htbp]
\begin{center}
\includegraphics[width=8cm]{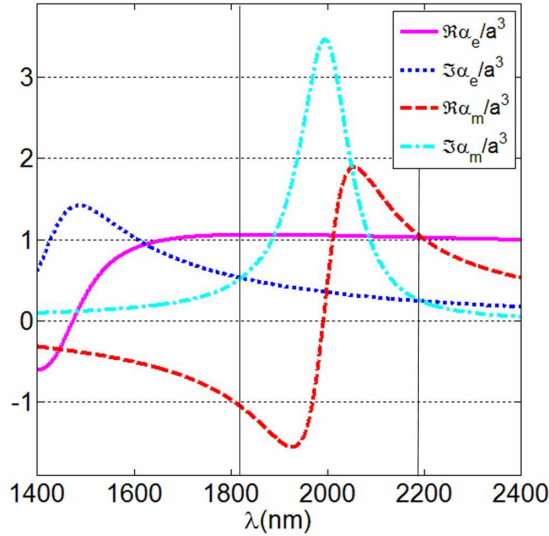}
\caption{Normalized real and imaginary parts of both the
electric and magnetic polarizabilities for a Ge sphere of radius $a=240nm$;
$\epsilon_p=16$ and $\mu_p=1$. The host medium has $\epsilon=\mu=1$. The first and second Kerker conditions are marked by the right and
left vertical lines, respectively.}
\label{GeAlpha}
\end{center}
\end{figure}

Figure \ref{GeAlpha} shows the real and imaginary parts of the
polarizabilities (Eq. \ref{Draine}) for a Ge sphere with $a=240$nm.
 One sees the values of
$\lambda$ at which
$\Im\{\alpha_e\}=\Im\{\alpha_m\}$, which are where the first and
second Kerker conditions hold for these polarizabilities.

\begin{figure}[htbp]
\begin{center}
\includegraphics[width=15cm]{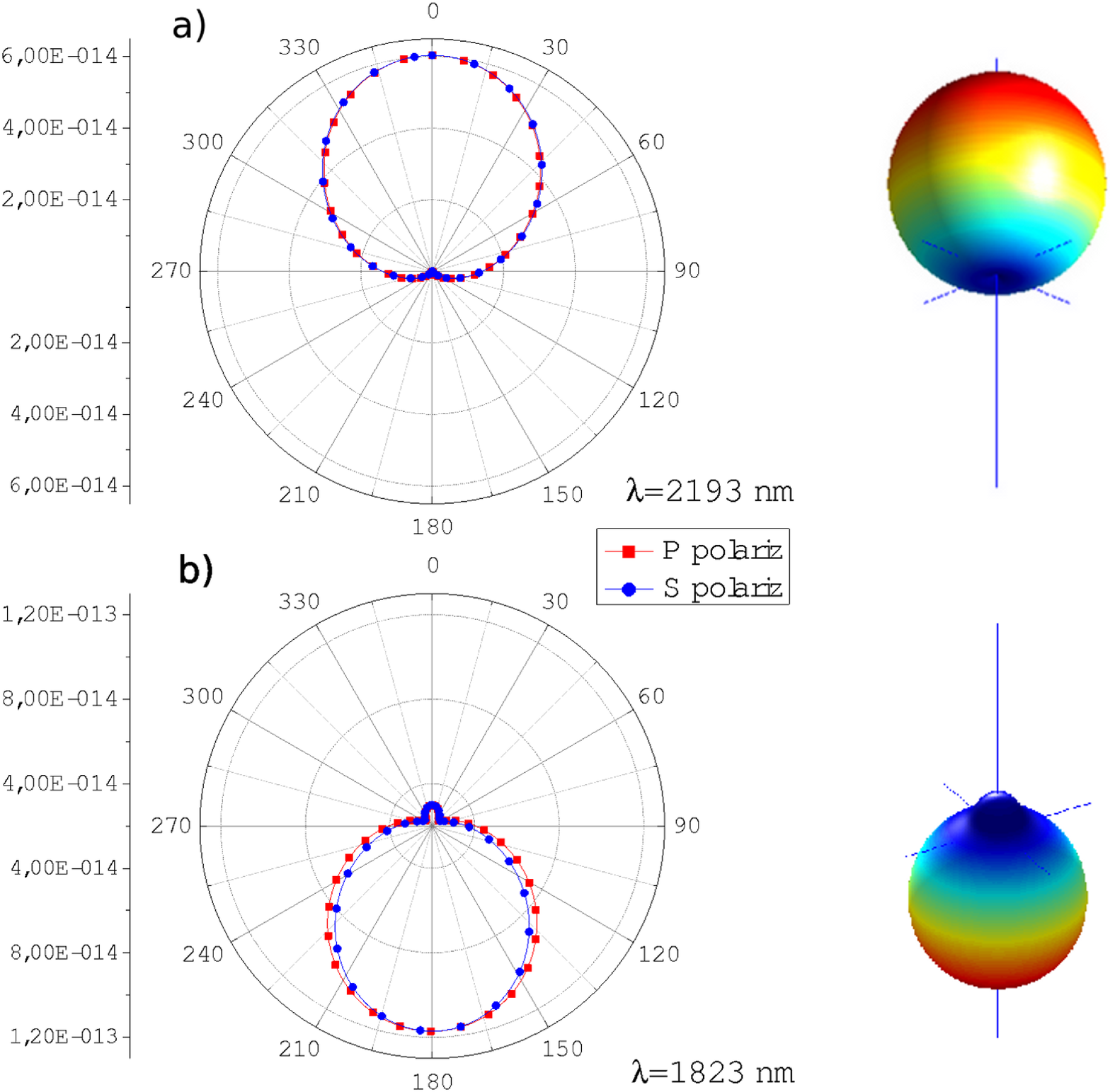}
\caption{Scattering diagrams for the 240nm Ge nanoparticle of Fig. \ref{GeMie}.
Both polarizations, with the incident electric field parallel (TM or P polarization) or normal (TE or S polarization) to the scattering plane are considered.}
\label{GeScatt}
\end{center}
\end{figure}

In order to check the validity of the dipolar approach and confirm the preditions at the Kerker condition, we have computed the exact scattering diagram from the full Mie expansion. The numerical results are shown in Fig. \ref{GeScatt}.
While the backward intensity drops to zero at the
first Kerker condition wavelength, at  the second
condition, although the most of the intensity goes backward, the scattering diagram presents a very small peak  in the
forward direction. As expected from the extinction cross section, the far field pattern is fully consistent with the dipolar approximation. In particular, at the Kerker's conditions it is independent of the incoming polarization. {\it Dielectric spheres and, in particular,
lossless Ge particles in the near infrared,  then constitute a
realizable laboratory to observe such a non-conventional scattering}.

\section{EFFECTS ON OPTICAL FORCES}

Let us now discuss some of the consequences of the Kerker's condition on optical forces.
The theory of the
force on a dipolar magnetic particle has recently  been developed
\cite{Chaumet_magnet,opex2010}. This
includes pure dielectric particles which can be  well described by
its first two electric and magnetic Mie coefficients \cite{opex2010}.
 The
time averaged force on a dipolar particle can be written as the sum of
three terms \cite{opex2010}:
\begin{eqnarray}
<{\bf F}> &=& <{\bf F}_{e}> + <{\bf F}_{m}>+ <{\bf F}_{e-m}> \nonumber
\\ &=& {\bf u}_z F_0  \left[ \frac{1}{a^3}\Im\left\{ \epsilon^{-1} \alpha_e  + \mu \alpha_m \right\}
- \frac{2k^3}{3a^3}\frac{\mu}{\epsilon}\Re(\alpha_e
\alpha_m^{\ast})\right].  \nonumber \\ \label{fpwsp1}
\end{eqnarray}
where $ F_0 \equiv \epsilon k a^3 |E_0|^2/2  $. The first two terms,
 $<{\bf F}_{e}>$ and $<{\bf F}_{m}>$, correspond to the forces on the induced pure electric and magnetic dipoles,
  respectively. $<{\bf F}_{e-m}>$, due to the interaction between both dipoles \cite{Chaumet_magnet,opex2010}, is
  related to the asymmetry in the scattered intensity distribution, (cf. the last term in Eq. \ref{difcross2})
  \cite{opex2010}.

  \begin{figure}[htbp]
\begin{center}
\includegraphics[width=8cm]{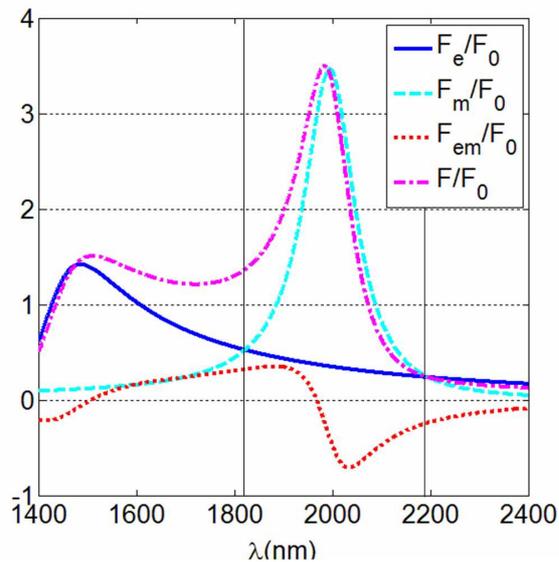}
\caption{Different contributions to the total radiation pressure,
versus the wavelenght, for the Ge particle of  Fig. \ref{GeMie}. Normalization
is done by $F_0=k\epsilon a^3|E_0|^2/2$ . Again, the vertical lines mark,
from right to left, the first and second Kerker conditions. Notice
that when the first Kerker condition is fullfilled, i.e.
$\Im{\alpha_e}=\Im{\alpha_m}$ and $\Re{\alpha_e}=\Re{\alpha_m}$,
$\langle F \rangle = \langle F_e \rangle = \langle F_m \rangle  = -
\langle F_{e-m} \rangle $. } \label{GeF}
\end{center}
\end{figure}

At the first generalized Kerker condition, Eq. \ref{firstKK}, the
interference term of Eq. \ref{fpwsp1} cancels out the magnetic
contribution and we obtain $<{\bf F}> =  <{\bf F}_e>$. At the second
Kerker condition, Eq. \ref{secondKK}, where the backscattering is
enhanced, $<{\bf F}> =  3 <{\bf F}_e>$. {\em Notice  that at both
Kerker conditions the scattering cross section is exactly the same;
however, the radiation pressures differ by a factor of 3}. These
properties are illustrated in Fig. \ref{GeF}, where we show the different
contributions to the total time averaged force on a submicron Ge
particle.

The strong peak in the radiation pressure force is mainly dominated by the first ``magnetic'' Mie resonance.
This is striking and in contrast with
all previous beliefs about optical forces on dipolar dielectric
particles, that assumed that these forces would solely be described
by the electric polarizability. It is also common to assume that for dielectric particles
the real part of the polarizability is much larger than its imaginary part. As a matter of fact, this is behind the development of optical tweezers, in which
gradient forces, (that are proportional to $\Re\{\alpha_e\}$), dominate over the radiation pressure or scattering force contribution, (which is proportional to
$\Im\{\alpha_e\}$) \cite{Quidant}. However, as the size of the particle increases, and {\em for any dielectric particle},  there is a crossover from electric to magnetic response as we
approach  the first Mie resonance, point at which the response is absolutely dominated by the magnetic dipole. Moreover,
 just at the resonance, and in absence of absorption,  $\Re\{\alpha_m\} =0$ and $\Im\{\alpha_m\} = 3/(\mu 2k^3)$.
 Then, the radiation pressure contribution of the magnetic term dominates the total force
 $<{\bf F}> \simeq <{\bf F}_m> \approx  (\epsilon |E_0|^2/2) [3/(2k^2)] $. Namely,
 {\em in resonance the radiation pressure force presents a strong peak,  the maximum force being independent of both
 material parameters and particle radius.}

\section{CONCLUSIONS}

In summary, we have predicted  that real small dielectric  particles
made of non-magnetic materials present non-conventional scattering
properties similar to those previously reported for somewhat
hypothetical  magnetodielectric particles \cite{Kerker}, resulting
from an interplay between real and imaginary parts of both electric
and magnetic polarizabilities. The exact scattering diagram, computed from the full Mie expansion, of submicron Ge particles in the infrared was shown to be consistent with the expected result for dipolar electric and magnetic scattering. Then we showed that these unusual
scattering effects do also affect the radiation pressure on these
small particles; specifically, the generalized Kerker's conditions have been tested on Ge spheres.
Submicron Ge
particles constitute an excellent laboratory to observe such new
scattering phenomena and force effects.
Being Ge permittivity higher than Si, the present work extend the range of some
possible applications (previously suggested for silicon particles)  to systems where the host medium presents
refractive index larger than vacuum.

The extraordinary scattering properties discussed here,
will strongly affect the dynamics of particle confinement in optical
traps and vortex lattices \cite{nanolet}  governed by both gradient and curl forces
\cite{opex2010,Albaladejo}. The interference between electric and magnetic dipoles suggest also intriguing possibilities regarding resonant coupling between the scattered dipolar field and guided modes in confined geometries
\cite{rgmprl1,rgmprl2}. We do believe that
 our results will stimulate further experimental and
theoretical work in different directions,  from optical
trapping and particle manipulation to cloaking and the design of
optical metamaterials based on lossless dielectric particles.


\section*{Acknowledgement}
We appreciate interesting discussions with  P. Albella, L.S. Froufe-P\'erez and  A. Garc\'ia-Etxarri.
 This work has been
supported by the  EU  NMP3-SL-2008-214107-Nanomagma,  the
Spanish MICINN Consolider \textit{NanoLight} (CSD2007-00046), FIS2009-13430-C01-C02 and FIS2007-60158,
as well as by the Comunidad de Madrid Microseres-CM (S2009/TIC-1476). Work by R.G.-M.  was supported by the MICINN ``Juan de la Cierva'' Fellowship.


\end{document}